\documentclass{article}

\usepackage[eandd,final]{neurips_2026}

\usepackage[utf8]{inputenc}
\usepackage[T1]{fontenc}
\usepackage{url}
\usepackage{booktabs}
\usepackage{amsfonts}
\usepackage{amsmath}
\usepackage{nicefrac}
\usepackage{microtype}
\usepackage{xcolor}
\usepackage{graphicx}
\usepackage{multirow}
\usepackage{enumitem}
\usepackage{algorithm}
\usepackage{algpseudocode}
\usepackage[hidelinks]{hyperref}

\newcommand{\cmark}{Yes}
\newcommand{\xmark}{No}

\newcommand{\makecell}[2][c]{\begin{tabular}{@{}#1@{}}#2\end{tabular}}
\makeatletter
\newcommand{\phoenixcaptionof}[1]{\def\@captype{#1}\caption}
\makeatother
\newcommand{\finding}[1]{\par\noindent\textbf{Main finding:}~\emph{#1}\par}

\title{Is Agentic AI Ready for Real-World Hardware Engineering? A Deep Dive with Phoenix-bench}

\author{%
  \textbf{Qingyun Zou} \\
  {\normalfont\small National University of Singapore} \\
  {\normalfont\small\href{mailto:qingyunzou@u.nus.edu}{qingyunzou@u.nus.edu}}
  \And
  \textbf{Feng Yu} \\
  {\normalfont\small National University of Singapore} \\
  {\normalfont\small\href{mailto:yuf@u.nus.edu}{yuf@u.nus.edu}}
  \And
  \textbf{Hongshi Tan} \\
  {\normalfont\small National University of Singapore} \\
  {\normalfont\small\href{mailto:hongshi@u.nus.edu}{hongshi@u.nus.edu}}
  \AND
  \textbf{Jiahao Cui} \\
  {\normalfont\small National University of Singapore} \\
  {\normalfont\small\href{mailto:cuijiahao@u.nus.edu}{cuijiahao@u.nus.edu}}
  \And
  \textbf{Bingsheng He} \\
  {\normalfont\small National University of Singapore} \\
  {\normalfont\small\href{mailto:dcsheb@nus.edu.sg}{dcsheb@nus.edu.sg}}
  \And
  \textbf{Weng-Fai Wong} \\
  {\normalfont\small National University of Singapore} \\
  {\normalfont\small\href{mailto:wongwf@nus.edu.sg}{wongwf@nus.edu.sg}}
}

\begin{document}

\maketitle

\begin{abstract}
We ask whether agentic artificial intelligence (AI) systems built for software engineering transfer to realistic hardware engineering. Existing benchmarks for hardware-oriented large language models (LLMs) isolate sub-tasks, but none jointly requires repository navigation, hierarchy-aware localization, executable verification with electronic design automation (EDA) tools, and maintenance-style patching. We introduce \textbf{Phoenix-bench}, a corpus of 511 verified Verilator instances from 114 GitHub repositories, each shipped with the developer patch, design-flow labels, fail-to-pass and pass-to-pass testbenches, and a shared open-source EDA toolchain. Using Phoenix-bench, we evaluate four commercial agents and eight open-source agentic structures across four LLM backbones, together with two diagnostic interventions: file-level oracle localization and testbench-log feedback. Agents that perform well on SWE-bench Verified degrade sharply on Phoenix-bench, and our analysis traces this gap to how hardware bugs propagate through module hierarchies, pointing to verification feedback as a key lever for hardware-aware agents.
\end{abstract}

\section{Introduction}

Agentic AI has been remarkably successful on software engineering tasks, where coding agents such as SWE-Agent~\citep{yang2024swe}, OpenHands~\citep{wang2024openhands}, mini-SWE-agent~\citep{miniswe2025}, and Agentless~\citep{xia2024agentless} are now capable of resolving a large fraction of real GitHub software repository issues end-to-end on benchmarks like SWE-bench~\citep{jimenez2023swe}. Despite the use of high-level programming languages, hardware design is fundamentally different: a hardware design is a circuit, i.e., a parallel network of instantiated modules wired together by signal flow, rather than a sequential program executed along a call graph.  Whether agents tuned for the call-graph world of software transfer to this parallel, signal-flow world of hardware is an open question. Resolving repository-level hardware issues raises a concrete question: will the same software coding agent be able to obtain hardware designs expressed in high-level languages and pass EDA tests, and, if not, what more is required?

Existing hardware LLM benchmarks isolate useful capabilities but do not yet cover the full hardware design workflow. Verilog generation benchmarks test whether a model can write standalone modules~\citep{thakur2024verigen,liu2023verilogeval,lu2024rtllm}; repair and debugging benchmarks test localized fixes~\citep{tsai2024rtlfixer,chen2024hdldebugger,ahmad2023fixing}; other work studies functional bug localization~\citep{yao2025location}, assertion-failure repair~\citep{zhou2025insights}, and assertion generation~\citep{fang2024assertllm}. These are valuable, but a project-wide full intervention requires repository navigation, hierarchy-aware localization, executable verification, and maintenance-style patching. It is unclear if existing software coding agents can perform this end-to-end. 

\begin{figure}[t]
  \centering
  \includegraphics[width=0.8\linewidth]{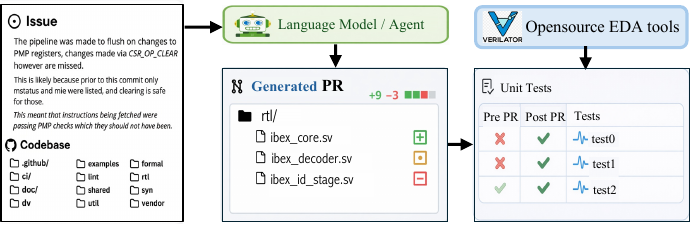}
  \caption{Phoenix-bench task overview: an agent edits a Verilog/SystemVerilog repository in response to a real GitHub issue and is judged by EDA fail-to-pass and pass-to-pass tests. PR: pull request.}
  \label{fig:repo_issue_example}
  \vspace{-0.5em}
\end{figure}

We introduce \textbf{Phoenix-bench}.\footnote{Code: \url{https://github.com/zanehpn/Phoenix-Bench}; benchmark data: \url{https://huggingface.co/datasets/OvertureMMXXI/Phoenix-Bench}.}
As shown in Figure~\ref{fig:repo_issue_example}, a benchmark in Phoenix-bench comprises a real GitHub issue and a full Verilog/SystemVerilog repository, and edits the repository so that the edited version passes both deterministic fail-to-pass and pass-to-pass Verilator tests. The synchronized corpus contains 511 verified instances from 114 GitHub repositories, selected from a crawl of 18,010 issues across 786 repositories, each shipped with developer patch, design-flow labels, and a shared open-source EDA toolchain so that resolved-rate differences reflect the agent's behavior rather than the availability of the tool.

Using Phoenix-bench, we did a comprehensive evaluation of four commercial coding agents and eight open-source agentic structures across four LLM backbones, along with two diagnostic interventions (file-level oracle localization and one round of testbench-log feedback). Three key findings emerge from our study: (i) Software and hardware are fundamentally different engineering tasks, and software coding agents do not perform well. For example, OpenHands with Qwen3-Coder-480B went from 69.6\% accuracy on SWE-bench Verified to  32.3\% on Phoenix-bench. 
(ii) The agents' failures concentrated in specific sets of issues not found in software, and on a handful of particularly difficult cases.
(iii) Localization granularity matters more than localization itself. A perfect file-level oracle yields only $1.4$\% improvement because the agent does not know what kinds of hardware bugs (synthesis, functions or others) and ignore files that were identified.

The contributions of this paper are:
\begin{itemize}[leftmargin=*, itemsep=2pt, topsep=2pt]
  \item \textbf{Phoenix-bench, a hardware-issue-resolution benchmark.} A synchronized corpus of 511 verified Verilator instances from 114 GitHub repositories, paired with developer patches, design-flow labels, fail-to-pass and pass-to-pass testbenches, and a shared open-source EDA toolchain so the resolved-rate metric reflects agent behavior rather than toolchain availability.
  \item \textbf{Comprehensive evaluation of commercial and open-source agents.} Four state-of-the-art commercial software coding agents (Claude Code, OpenAI Codex, Gemini CLI, GitHub Copilot) and eight open-source agentic structures across four LLM backbones, plus two diagnostic interventions (file-level oracle localization and testbench-log feedback) were tested using Phoenix-bench.
  \item \textbf{Actionable hardware-specific findings for future AI approaches.} An analysis of the evaluation results yielded insights that we believe will improve the design of future agents for hardware design. 
\end{itemize}
\section{Related Work: Hardware Code Benchmarks}
\label{sec:related}


Existing hardware code benchmarks are usually described by visible attributes such as module versus repository, generation versus repair, and exact-match versus simulation. These attributes are useful but insufficient: they do not explain why a benchmark reflects, or fails to reflect, real hardware engineering. We instead ask whether a benchmark captures repository context, cross-module signal dependencies, execution-grounded EDA verification, maintenance-style patching, and issue diversity across the hardware design flow. Table~\ref{tab:dataset_comparison} separates hardware benchmarks by the engineering behavior they require. Module-level benchmarks such as VerilogEval~\citep{liu2023verilogeval}, RTLLM~\citep{lu2024rtllm}, HDLEval~\citep{kashanaki2024hdleval}, PyHDL-Eval~\citep{batten2024pyhdl}, RTLFixer~\citep{tsai2024rtlfixer}, and HDLdebugger~\citep{chen2024hdldebugger} test generation, repair, or debugging in restricted contexts, so they do not require cross-file consistency or hierarchy-level signal tracing. Repository-level completion benchmarks such as RTL-Repo~\citep{allam2024rtlrepo} and MHRC-Bench~\citep{zou2026mhrc} evaluate code completion with repository context. Their completion metrics differ from issue-resolution success; MHRC-Bench also reports compilation pass rates. HWFixBench~\citep{fu2025hwfixbench} evaluates hardware bug fixing and patch generation with an LLM judge and also provides a simulator backend. Phoenix-bench evaluates repository-level issue resolution through executable EDA verification.

Register-transfer level (RTL) generation also benefits from specialized datasets and execution-based training signals. EARL~\citep{shi2026earl} aligns Verilog generators with verifiable rewards and entropy-guided policy updates focused on high-uncertainty tokens. HLStrans~\citep{zou2025hlstrans} provides paired C and high-level synthesis (HLS) programs with testbenches and synthesis-derived latency and resource annotations. HLS-Seek~\citep{zou2026hlsseek} trains HLS code generators with comparative proxy rewards for latency and resource use, invoking synthesis selectively when proxy uncertainty is high. Reward-weighted on-policy distillation from natural language (NL) to SystemVerilog Assertions (SVA)~\citep{zou2026rwopd} uses an open property-equivalence verifier to select and weight student rollouts. These studies address RTL generation, C-to-HLS synthesis, and assertion generation; Phoenix-bench evaluates repository-level issue resolution.

\begin{table*}[t]
\centering
\caption{Coverage of hardware engineering practice across existing hardware benchmarks. Testbench repair denotes tasks requiring changes to verification code, rather than the availability of evaluation testbenches. EM: exact match; ES: edit similarity.}
\label{tab:dataset_comparison}
\scriptsize
\setlength{\tabcolsep}{2.5pt}
\renewcommand{\arraystretch}{1.15}
\resizebox{\textwidth}{!}{%
\begin{tabular}{l l l l l l}
\toprule
\textbf{Benchmark}
& \textbf{\makecell{Engineering\\setting}}
& \textbf{Task}
& \textbf{Verification}
& \textbf{\makecell{Design-flow\\coverage}}
& \textbf{\makecell{Testbench\\repair}} \\
\midrule
VerilogEval~\citep{liu2023verilogeval}
& Isolated module
& Code generation
& EDA simulation
& Design only
& \xmark \\
RTLLM~\citep{lu2024rtllm}
& Isolated module
& Code generation
& EDA simulation
& Design only
& \xmark \\
RTLFixer~\citep{tsai2024rtlfixer}
& Isolated module
& Syntax repair
& Compilation
& Design only
& \xmark \\
HDLdebugger~\citep{chen2024hdldebugger}
& Isolated module
& Debug repair
& Simulation
& Design/verification
& \xmark \\
\midrule
RTL-Repo~\citep{allam2024rtlrepo}
& Repository snippets
& Completion
& EM / ES
& Design only
& \xmark \\
MHRC-Bench~\citep{zou2026mhrc}
& Repository context
& Completion
& \makecell{EM / ES / CodeBLEU\\Compilation}
& Design only
& \xmark \\
HWFixBench~\citep{fu2025hwfixbench}
& Repository repair
& Patch generation
& LLM judge
& Limited
& \xmark \\
\textbf{Phoenix-bench}
& \textbf{Real issue + repo}
& \textbf{Bug repair}
& \textbf{EDA execution}
& \textbf{Different stages}
& \textbf{\cmark} \\
\bottomrule
\end{tabular}
}
\end{table*}

\section{Phoenix-bench Benchmark Suite}
\label{sec:benchmark}

This section describes the Phoenix-bench workload, including data collection, issue categories, artifact statistics, and verification.

\textbf{Design principles.} Phoenix-bench uses five design principles to keep the task close to hardware maintenance. First, each issue is resolved in the full repository, because the relevant behavior may span instantiated modules, packages, generated artifacts, and build scripts. Second, localization must be able to follow hierarchical signal propagation instead of relying only on lexical overlap. Third, correctness is measured by simulator, compiler, synthesis, or exact artifact checks. Fourth, issue categories cover design, verification, toolchain, implementation, and documentation artifacts. Fifth, the output is a repository patch, not standalone RTL.

\begin{figure*}[t]
  \centering
  \includegraphics[width=0.9\textwidth]{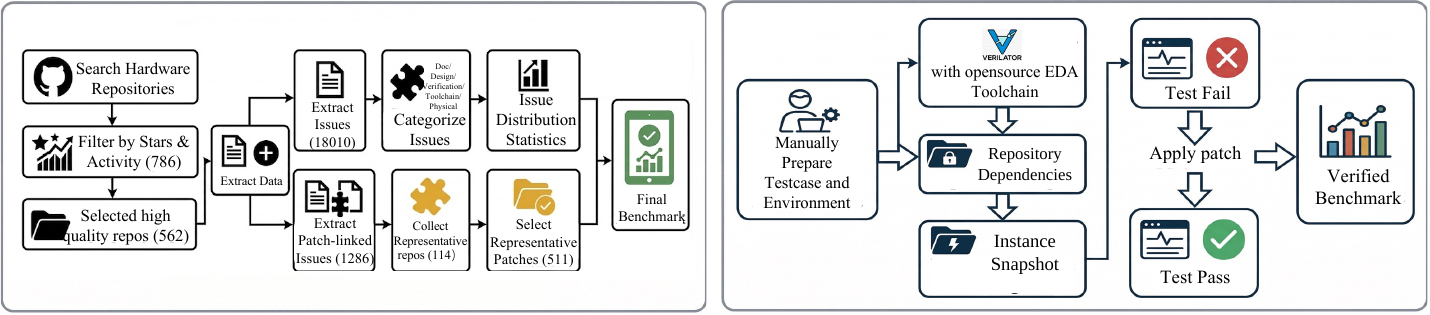}
  \caption{Phoenix-bench construction pipeline, from GitHub crawl to verified benchmark instances.}
  \label{fig:benchmark_construction}
\end{figure*}

\textbf{Data collection.} Phoenix-bench is built through the four-stage funnel in Figure~\ref{fig:benchmark_construction}. We first query the GitHub Search application programming interface (API) for Verilog/SystemVerilog repositories with at least 50 stars, then retain issues linked to pull requests (PRs) through explicit references or GitHub native linking. We next filter for PRs that modify Verilog/SystemVerilog artifacts and remove duplicate or incomplete cases. This process yields a synchronized runner corpus of 511 verified Verilator instances with both fail-to-pass and pass-to-pass tests; category, runner, repository, and coverage statistics all use this same 511-instance corpus.

\textbf{Design-flow categories.} Phoenix-bench labels issues by the hardware artifact whose behavior must change. \emph{Design} issues cover RTL behavior such as functional bugs, specification mismatches, and quality-of-results concerns. \emph{Verification} issues cover testbench, assertion, or simulation-build failures. \emph{Environment \& Toolchain} issues cover simulator compatibility, build scripts, and tool invocation. \emph{Physical \& Implementation} issues cover synthesis-stage effects such as timing warnings, latch inference, and resource use. \emph{Documentation} issues cover README files, comments, and API descriptions. We stop at synthesis because backend stages such as place-and-route and clock-tree synthesis are not deterministically reproducible across the collected repositories in our current environment.

\begin{figure*}[t]
  \centering
  \begin{minipage}[t]{0.58\textwidth}
    \centering
    \includegraphics[width=\linewidth]{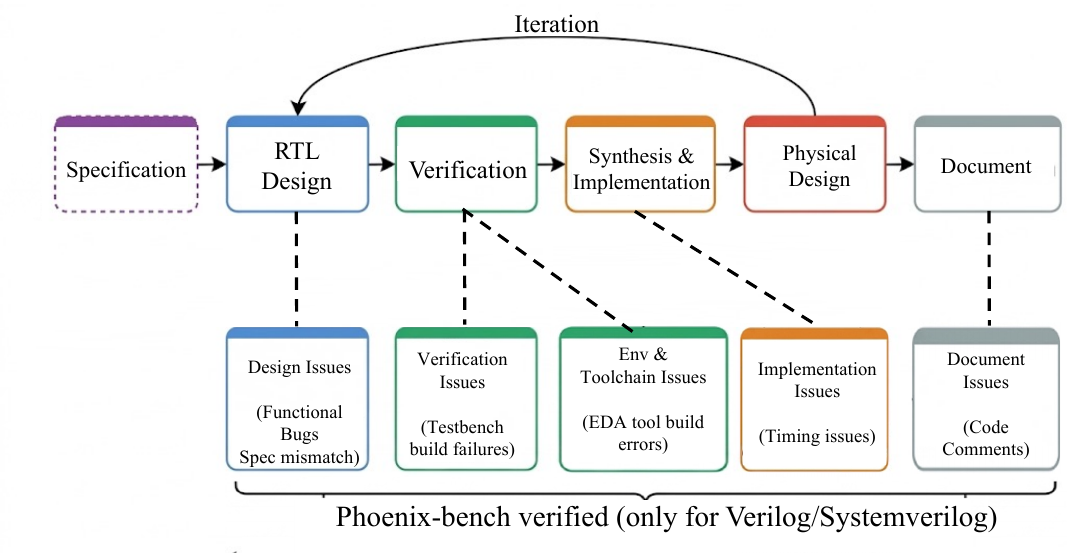}
    \small (a) ASIC/FPGA design flow
  \end{minipage}
  \hfill
  \begin{minipage}[t]{0.38\textwidth}
    \centering
    \includegraphics[width=\linewidth]{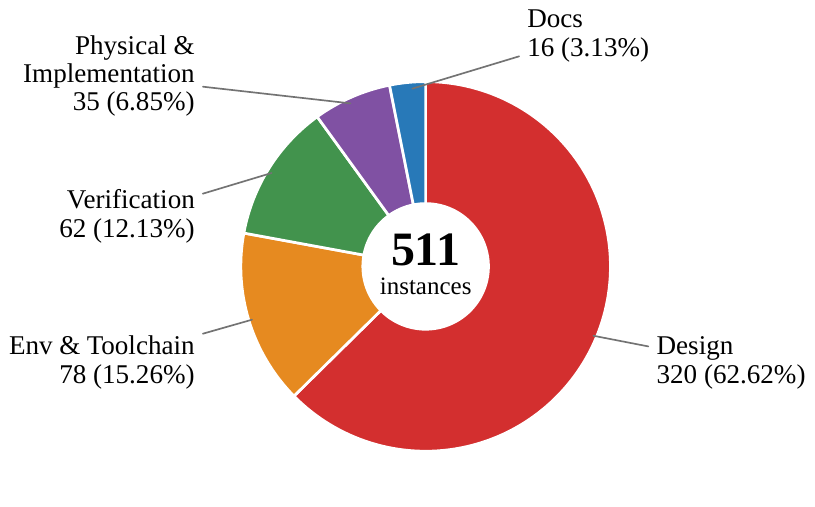}
    \small (b) Issue category distribution
  \end{minipage}
  \caption{Application-specific integrated circuit (ASIC) / field-programmable gate array (FPGA) design flow mapped to Phoenix-bench issue categories (left) and the 511-instance issue-category distribution (right).}
  \label{fig:categories_prof}
\end{figure*}

\begin{figure*}[t]
\centering
\begin{minipage}[t]{0.32\textwidth}
  \vspace{0pt}
  \centering
  \includegraphics[width=\linewidth]{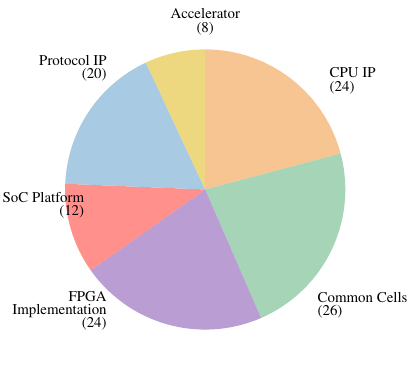}
  \phoenixcaptionof{figure}{Repository-application composition across the 114 GitHub repositories used in Phoenix-bench. CPU: central processing unit.}
  \label{fig:issue_stats}
\end{minipage}
\hfill
\begin{minipage}[t]{0.64\textwidth}
\vspace{0pt}
\centering
\phoenixcaptionof{table}{Benchmark statistics: SWE-bench Verified~\citep{openai_swebench_verified_2024} (500 Python instances) versus the current Phoenix-bench synchronized runner corpus (511 Verilog/SystemVerilog instances). TB: testbench.}
\label{tab:dataset_stats}
\resizebox{\linewidth}{!}{%
\begin{tabular}{l l c c c c}
\toprule
 & & \multicolumn{2}{c}{\textbf{SWE-bench}} & \multicolumn{2}{c}{\textbf{Phoenix-bench}} \\
 & & \textbf{Mean} & \textbf{Max} & \textbf{Mean/Agg.} & \textbf{Max} \\
\midrule
\textbf{Issue Text}
& Length (words)
& 195.1 & 4,477
& 171.7 & 2,104 \\
\midrule
\textbf{Codebase}
& \# Files
& 3,010 & 5,890
& 1,153.2 & 13,061 \\
& \# Lines
& 438K & 886K
& 132.7K & 2.08M \\
\midrule
\textbf{Gold Patch}
& \# Lines edited (HDL)
& 32.8 & 5,888
& 735.5 & 113,849 \\
& \# Files edited (HDL)
& 1.7 & 31
& 5.7 & 368 \\
& \# Patch scopes edited
& 3.0 & 36
& 8.4 & 348 \\
\midrule
\textbf{Tests}
& \# Fail-to-pass tests/checks
& 9.1 & 1,633
& 3.5 & 82 \\
& \# TB Lines
& 120.8 & 9,459
& 49.8 & 219 \\
& Patch Line Coverage (\%)
& -- & --
& 99.91 & 100 \\
& Patch Branch Coverage (\%)
& -- & --
& 100.0 & 100 \\
& Patch Toggle Coverage (\%)
& -- & --
& 100.0 & 100 \\
& Patch Function Coverage (\%)
& -- & --
& 100.0 & 100 \\
\bottomrule
\end{tabular}%
}
\end{minipage}
\end{figure*}

\textbf{Dataset composition.} The collected issues are not distributed like the final benchmark. Documentation issues dominate the raw crawl, but code-linked issues shift toward toolchain and design changes because those categories more often produce concrete PR patches. Figure~\ref{fig:categories_prof}(b) shows all 511 instances: 320 Design, 78 Environment and Toolchain, 62 Verification, 35 Physical and Implementation, and 16 Documentation issues, organized into 23 subcategories. Design issues account for 62.6\% of the corpus. These instances span 114 GitHub repositories, including processor cores, common-cell and protocol intellectual property (IP) libraries, FPGA/system-on-chip (SoC) applications, and simulator/toolchain repositories. Table~\ref{tab:dataset_stats} reports repository, patch, and testbench statistics together with patch-local coverage. Issue-text length counts whitespace-separated words in each complete released issue file, including metadata, appended PR content, and reproducer notes. The harness emits 71,048 log entries prefixed with \texttt{TESTCASE:} across the synchronized corpus. These entries include patch-derived file, hunk, and line summaries, execution-status messages, and coverage reports, so their total does not measure the number of independent functional tests or assertions. Patch coverage is computed over patch-touched executable lines, Verilator branch/toggle points, and recovered patch-local function or scope entries, which keeps the coverage metric tied to the code changed by the developer patch.

\textbf{Verification environment.} Each instance is verified with a portfolio of open-source EDA tools: Verilator for cycle-accurate simulation, Icarus Verilog as an alternative event-driven simulator for projects whose testbenches use SystemVerilog constructs Verilator does not support, Surelog for SystemVerilog parsing and elaboration, and Yosys for synthesis and lint-style checks. A shared toolchain provides these tools, per-repository setup supplies project dependencies, and each instance pins the issue snapshot. Two execution checks gate acceptance: a \emph{fail-to-pass} check requires the targeted testbench to fail on the buggy snapshot and pass after applying the developer patch, and a \emph{pass-to-pass} check requires the agent's patch to keep all previously-passing testbenches still passing so the patch is not allowed to break unrelated functionality. All 511 instances are verified through this open-source EDA toolchain, and the acceptance criterion for all headline results is execution-grounded verification, not LLM judgment.

\section{Experiments: Product Agents, Open-source Agents, and Diagnostics}
\label{sec:eval}

\subsection{Evaluation protocol}

We use Phoenix-bench to study three questions. \textbf{RQ1}: How well do real product coding agents handle repository-level hardware tasks under EDA verification? (\S\ref{sec:results}, Table~\ref{tab:product_results}) \textbf{RQ2}: Which open-source agent structures from software engineering transfer to hardware engineering? (\S\ref{sec:results}, Table~\ref{tab:agentic_study}) \textbf{RQ3}: Where do current failures concentrate, and which factors, task domain, patch complexity, localization, or testbench feedback, best explain the residual? (decomposed in \S\ref{sec:hdl_taxonomy} to \S\ref{sec:eda_feedback})

\textbf{Evaluation modes.} Phoenix-bench separates external validity from mechanism control. The \emph{product-agent} mode evaluates deployed coding agents as black boxes through their native interfaces, preserving their orchestration, model selection, planning, and editing behavior. The \emph{open-source-agent} mode runs reproducible open-source agent structures under the same EDA evaluator and a shared set of LLM backbones, so that model choice, localization strategy, and testbench-feedback handling can be compared directly.

\textbf{Product agents.} We evaluate four state-of-the-art commercial coding agents through their native interfaces, namely Claude Code~\citep{anthropic_claude_code_2026}, OpenAI Codex~\citep{openai_codex_2026}, Gemini CLI~\citep{google_gemini_cli_2026}, and GitHub Copilot coding agent~\citep{github_copilot_coding_agent_2026}. These systems are evaluated as black boxes because their engineering value lies in the complete product loop, not only in the underlying model. Each product receives the same issue text, repository snapshot, EDA verifier, time budget, and final acceptance gate. We record the generated patch, transcript or action log when available, EDA output, wall-clock time, test-call count, and token or credit cost when the product exposes it. \textbf{Open-source agents.} We additionally instantiate or adapt eight representative agentic structures under Phoenix-bench in three categories. \emph{Agent-based}: mini-SWE-agent~\citep{miniswe2025}, SWE-agent~\citep{yang2024swe}, OpenHands~\citep{wang2024openhands}, Aider~\citep{aider2024}, KGCompass~\citep{kgcompass2025}, and Lingxi~\citep{lingxi2025}. \emph{Retrieval-augmented}: ExpeRepair~\citep{mu2025experepair}. \emph{Workflow-based}: Agentless~\citep{xia2024agentless}. All open-source agents are given the same hardware adaptations where applicable, including EDA toolchain access and SystemVerilog-aware filtering.


\textbf{Models.} Each product agent is evaluated with its default backbone; the specific model used at the time of writing is reported in the Backbone column of Table~\ref{tab:product_results}. The open-source agents are evaluated with four LLM backbones spanning proprietary and open-source systems: GPT-5.2~\citep{openai_5_2_system_card}, Gemini-3-Pro~\citep{google_gemini_3_pro}, DeepSeek-V3.2~\citep{liu2025deepseek}, and Qwen3-Coder-480B~\citep{qwen3coder2025}. Additional model results can be integrated into the same protocol.

\textbf{Metrics.} \emph{Resolved Rate} is the primary end-to-end metric: a task is resolved only if the generated patch passes the EDA testbench execution. \emph{File Precision} is the fraction of agent-modified files that appear in the developer patch, and \emph{File Recall} is the fraction of developer-patch files modified by the agent. \emph{Module Precision} and \emph{Module Recall} are computed by extracting \texttt{module}/\texttt{endmodule} boundaries from both patches and mapping each modified region to its enclosing module. Product-agent reports additionally include wall-clock time, cost or credit usage when exposed, and EDA-test invocation counts. For benchmark reliability, leaderboard submissions should report headline resolved rates with bootstrap confidence intervals over instances.

\begin{table*}[!t]
\centering
\caption{Product-agent results on Phoenix-bench under the same EDA verifier. \textbf{Bold} marks the best value per column.}
\label{tab:product_results}
\small
\setlength{\tabcolsep}{4pt}
\resizebox{\textwidth}{!}{%
\begin{tabular}{l l rrrrrr}
\toprule
\textbf{Product agent} & \textbf{Backbone} & \makecell{Resolved\\Rate (\%)} & \makecell{File\\Prec. (\%)} & \makecell{File\\Recall (\%)} & \makecell{Module\\Prec. (\%)} & \makecell{Module\\Recall (\%)} & \makecell{Tokens\\(/inst.)} \\
\midrule
Claude Code                  & Claude Opus 4.7   & \textbf{38.6} & 38.6          & \textbf{54.2} & 47.9          & \textbf{48.3} & 432k \\
OpenAI Codex                 & GPT-5.5           & 38.0          & 41.2          & 52.5          & \textbf{49.7} & 47.1          & 326k \\
Gemini CLI                   & Gemini-3.1-Pro    & 37.4          & 37.4          & 50.9          & 46.5          & 45.4          & 254k \\
GitHub Copilot coding agent  & GPT-5.3-Codex     & 32.7          & \textbf{47.8} & 43.6          & 48.4          & 40.2          & 148k \\
\bottomrule
\end{tabular}%
}

\caption{Open-source agentic factor study on Phoenix-bench (\%). \textbf{Bold}/\underline{underline} mark the top-1/top-2 values per row.}
\label{tab:agentic_study}
\small
\setlength{\tabcolsep}{2pt}
\resizebox{\textwidth}{!}{%
\begin{tabular}{ll r r rrrrrr}
\toprule
\multirow{2}{*}{Model} & \multirow{2}{*}{Metric}
& {Retrieval}
& {Workflow}
& \multicolumn{6}{c}{Agent-based} \\
\cmidrule(lr){3-3} \cmidrule(lr){4-4} \cmidrule(lr){5-10}
& & ExpeRepair & Agentless & \makecell{mini-\\SWE} & SWE-Agent & OpenHands & Aider & KGCompass & Lingxi \\
\midrule
\multirow{5}{*}{GPT-5.2}
& Resolved Rate      & 11.7 & 18.8 & 14.5 & \underline{27.8} & \textbf{33.9} & 16.2 & 7.2 & 6.3 \\
& File Precision      & \underline{52.7} & 45.8 & 45.3 & \textbf{60.1} & 36.2 & 45.1 & 32.1 & 34.3 \\
& File Recall         & 24.2 & 37.5 & 32.2 & \underline{49.4} & \textbf{50.7} & 34.6 & 20.6 & 22.5 \\
& Module Precision    & \underline{50.2} & 37.7 & 40.1 & \textbf{53.0} & 46.3 & 38.6 & 26.4 & 28.3 \\
& Module Recall       & 24.0 & 32.0 & 31.1 & \underline{44.5} & \textbf{45.0} & 31.1 & 17.6 & 19.2 \\
\midrule
\multirow{5}{*}{Gemini-3-Pro}
& Resolved Rate      & 10.8 & 13.3 & 13.3 & \underline{27.0} & \textbf{33.5} & 15.3 & 7.2 & 6.3 \\
& File Precision      & \underline{54.3} & 36.9 & 42.0 & \textbf{59.5} & 35.6 & 44.6 & 25.8 & 27.6 \\
& File Recall         & 19.0 & 31.1 & 32.7 & \underline{49.3} & \textbf{49.8} & 34.5 & 17.1 & 18.7 \\
& Module Precision    & \textbf{51.3} & 28.7 & 34.5 & \underline{50.1} & 45.4 & 37.6 & 20.1 & 21.6 \\
& Module Recall       & 18.8 & 25.1 & 30.2 & \underline{43.1} & \textbf{44.2} & 30.2 & 13.8 & 15.1 \\
\midrule
\multirow{5}{*}{DeepSeek-V3.2}
& Resolved Rate      & 7.2 & 10.8 & 8.0 & \underline{16.2} & \textbf{26.0} & 9.0 & 4.5 & 3.5 \\
& File Precision      & \textbf{57.9} & 22.5 & 43.7 & \underline{48.4} & 36.6 & 36.3 & 15.7 & 16.9 \\
& File Recall         & 11.6 & 19.2 & 24.2 & \underline{39.8} & \textbf{49.2} & 27.8 & 10.6 & 11.5 \\
& Module Precision    & \textbf{55.4} & 17.1 & 42.1 & 39.5 & \underline{46.3} & 29.6 & 11.9 & 12.8 \\
& Module Recall       & 11.5 & 15.3 & 24.2 & \underline{34.5} & \textbf{42.9} & 24.2 & 8.4 & 9.1 \\
\midrule
\multirow{5}{*}{\makecell[l]{Qwen3-\\Coder-480B}}
& Resolved Rate      & 10.0 & 12.5 & 12.5 & \underline{24.3} & \textbf{32.3} & 13.3 & 6.3 & 5.5 \\
& File Precision      & \textbf{59.7} & 31.5 & 36.3 & \underline{53.8} & 34.3 & 40.3 & 22.0 & 23.6 \\
& File Recall         & 16.2 & 25.3 & 31.3 & \underline{47.4} & \textbf{49.6} & 33.2 & 13.9 & 15.2 \\
& Module Precision    & \textbf{57.9} & 25.1 & 30.5 & 44.3 & \underline{45.7} & 33.2 & 17.6 & 18.8 \\
& Module Recall       & 16.2 & 21.6 & 28.9 & \underline{41.2} & \textbf{42.4} & 28.9 & 11.9 & 13.0 \\
\bottomrule
\end{tabular}
}
\end{table*}
\vspace{-1em}
\subsection{Results}
\label{sec:results}

Table~\ref{tab:product_results} reports the head-to-head results for the four product agents on Phoenix-bench. Across resolved rate and localization recall, the agents differ not only in raw success rate but also in how aggressively they invoke the EDA toolchain and in their per-instance wall-clock and cost profile. Agents with deeper testbench-feedback loops resolve more cases at the price of larger token and time budgets, while issue-to-PR style agents finish faster but recover fewer developer-patch files.

Beyond resolved rate, the localization metrics show that file-level coverage saturates earlier than module-level coverage: agents that locate the right files still miss enclosing modules and connected ports, which is consistent with the hardware-specific localization story Phoenix-bench is designed to expose. The per-instance token column reports inference usage alongside these effectiveness metrics.

Table~\ref{tab:agentic_study} shows that software-oriented open-source agents transfer only partially to hardware repositories. Interactive systems such as OpenHands and SWE-agent resolve more cases than static retrieval or workflow systems, but their file-level exploration still misses hierarchy-coupled dependencies. KGCompass and Lingxi perform weakly in this setting, which suggests that software knowledge graphs and software-style multi-agent decomposition are not enough to model HDL signal flow or testbench feedback. Hardware-aware localization changes the precision-recall trade-off: conservative systems such as ExpeRepair keep high file precision but miss many developer-patch files, while broad interactive systems recover more files at lower precision. Stronger LLM backbones improve all systems, but the relative ordering between agents remains stable.

\begin{figure}[t]
  \centering
  \begin{minipage}[b]{0.45\linewidth}
    \centering
    \includegraphics[width=\linewidth]{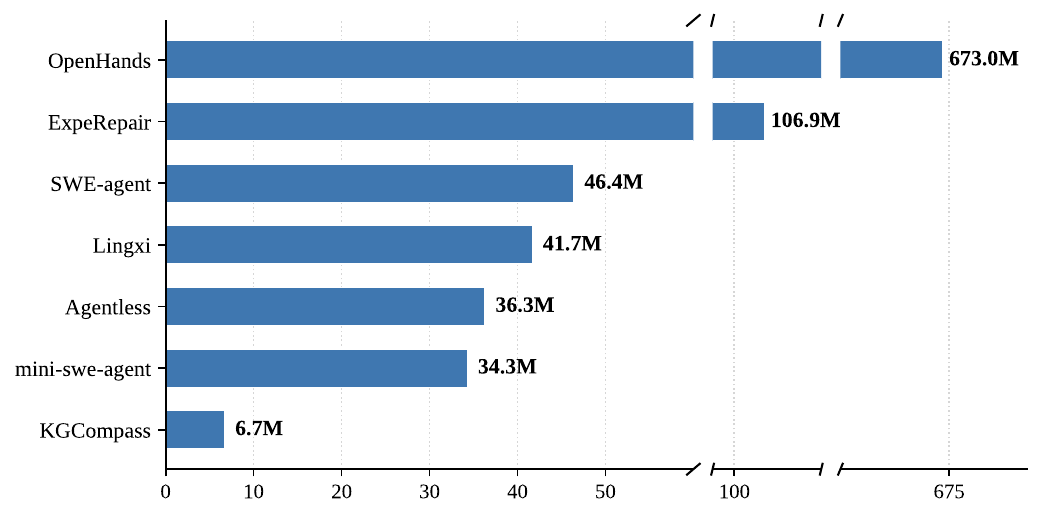}
    \phoenixcaptionof{figure}{Total token consumption across 511 instances for the seven open-source agents with collected token-usage data; Aider is excluded (broken x-axis).}
    \label{fig:token_usage}
  \end{minipage}\hfill
  \begin{minipage}[b]{0.51\linewidth}
    \centering
    \includegraphics[width=\linewidth]{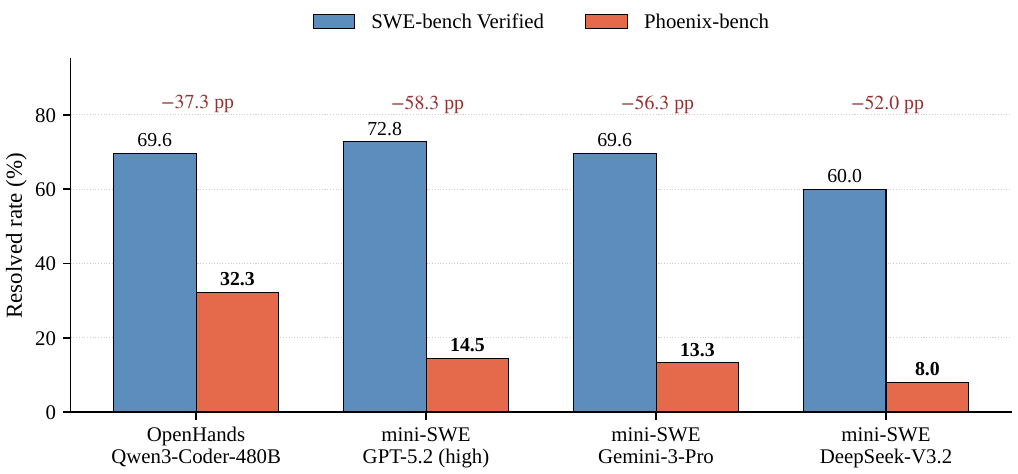}
    \phoenixcaptionof{figure}{SWE-bench Verified vs.\ Phoenix-bench for four representative (agent, backbone) pairs. pp: percentage points.}
    \label{fig:cross_benchmark}
  \end{minipage}
\end{figure}

Figure~\ref{fig:token_usage} reports total token consumption across the 511 Phoenix-bench instances for the seven open-source agents with collected token-usage data. Token-usage data were not collected for Aider. Token consumption varies substantially within the agent-based category: OpenHands uses 673.0M tokens, SWE-agent uses 46.4M, and KGCompass uses 6.7M. The workflow-based Agentless uses 36.3M tokens, while the retrieval-augmented ExpeRepair uses 106.9M. Together with the resolved rates and localization recall in Table~\ref{tab:agentic_study}, these totals show the resource use and task performance of each system and complement the per-instance token counts for product agents in Table~\ref{tab:product_results}.

\section{Key Insights}
\label{sec:analysis}

\subsection{How well do software agents fare on hardware design repositories?}
\label{sec:cross_benchmark}

Figure~\ref{fig:cross_benchmark} contrasts each system's public SWE-bench Verified score~\citep{swebench_leaderboard_2026} with its Phoenix-bench score for four representative (agent, backbone) pairs (rows in Table~\ref{tab:agentic_study}): all four sit in the $60$ to $73\%$ band on SWE-bench yet lose $37$ to $58$ percentage points on Phoenix-bench, the strongest SWE-bench pair (mini-SWE-agent + GPT-5.2 at $72.8\%$) scores $14.5\%$ on Phoenix-bench, while mini-SWE-agent + DeepSeek-V3.2 scores $8.0\%$, and the spread tracks the agent rather than the backbone---mini-SWE-agent drops $52$ to $58$ percentage points across GPT-5.2 / Gemini-3-Pro / DeepSeek-V3.2, whereas OpenHands + Qwen3-Coder-480B drops only $37$ percentage points despite a weaker backbone. The headline number also hides two qualitative shifts: SWE-bench failures are Python-style edge-case logic and missing tests~\citep{jimenez2023swe}, while Phoenix-bench failures concentrate in hardware-specific categories (port/width, clock/reset polarity, parameter/hierarchy, tool-pragma, procedural-construct misuse), and gold-patch shape changes accordingly---the median SWE-bench patch touches 1 file / 1 hunk / 7 lines with only $14.2\%$ multi-file, vs.\ the median Phoenix-bench patch at 3 files / 7 hunks / 63 edited lines, with $352/511$ ($68.9\%$) touching multiple distinct file paths across all file types. Interaction depth, not raw backbone strength, is therefore the agent-side property that most reliably survives the software-to-hardware shift.

HDLs describe circuits as many parallel instantiated modules wired together by signal flow, so fixing a bug typically requires following a signal across instantiations to find the related code. A specific behavioral mismatch in such a setting explains much of the per-agent spread: localization stops at the wrong granularity. On SWE-bench an agent reaches the offending function in one or two file-browse steps, but on Phoenix-bench the same browse pattern finds the file where the symptom appears but does not trace back through the instantiation chain. This surfaces later as the $77\%$ upstream-failure share in \S\ref{sec:hdl_taxonomy} and the $-113$ file-oracle regressions in \S\ref{sec:oracle}. It is this issue rather than a capability gap in the underlying LLM that explains the failures.

\finding{Software and hardware are fundamentally different engineering tasks: the same agent loses $37$ to $58$ percentage points from SWE-bench Verified to Phoenix-bench because hardware bugs propagate across parallel instantiated modules through signal flow rather than living in one file along a sequential call graph, so software-tuned agents stop at the symptom file rather than tracing back through the instantiation chain.}

\subsection{Where do unresolved failures concentrate?}
\label{sec:hdl_taxonomy}

\begin{figure}[t]
  \centering
  \includegraphics[width=\linewidth]{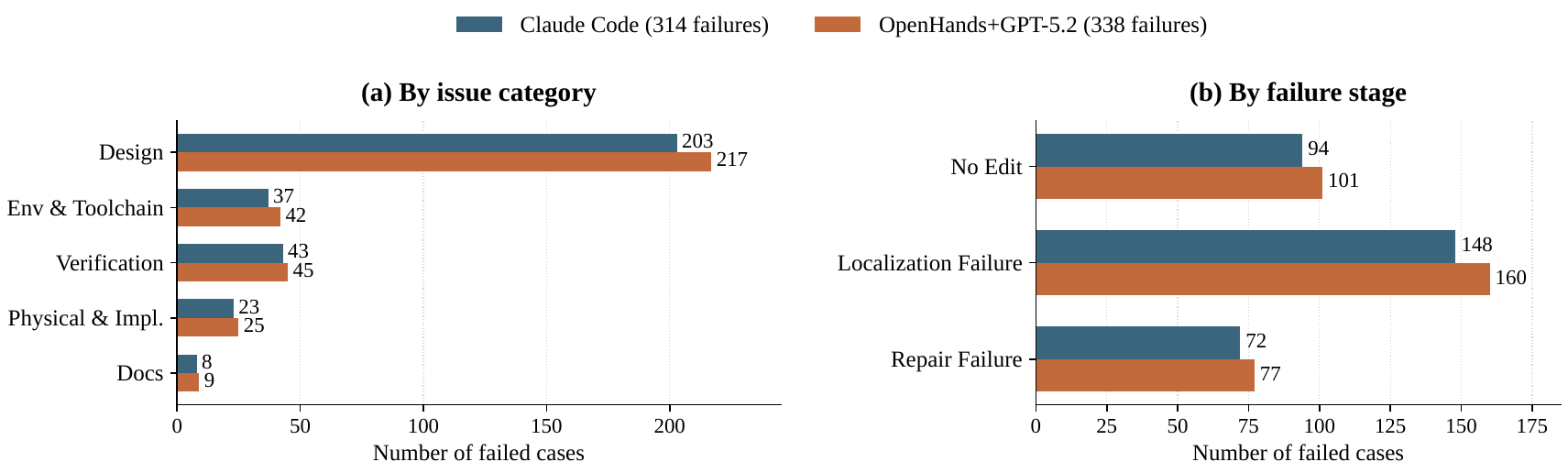}
  \caption{Failure distributions on the same 511 instances for Claude Code and OpenHands+GPT-5.2: (a) by issue category and (b) by failure stage, each summed over all failures.}
  \label{fig:failure_category}
\end{figure}

\begin{figure*}[t]
  \centering
  \includegraphics[width=\textwidth]{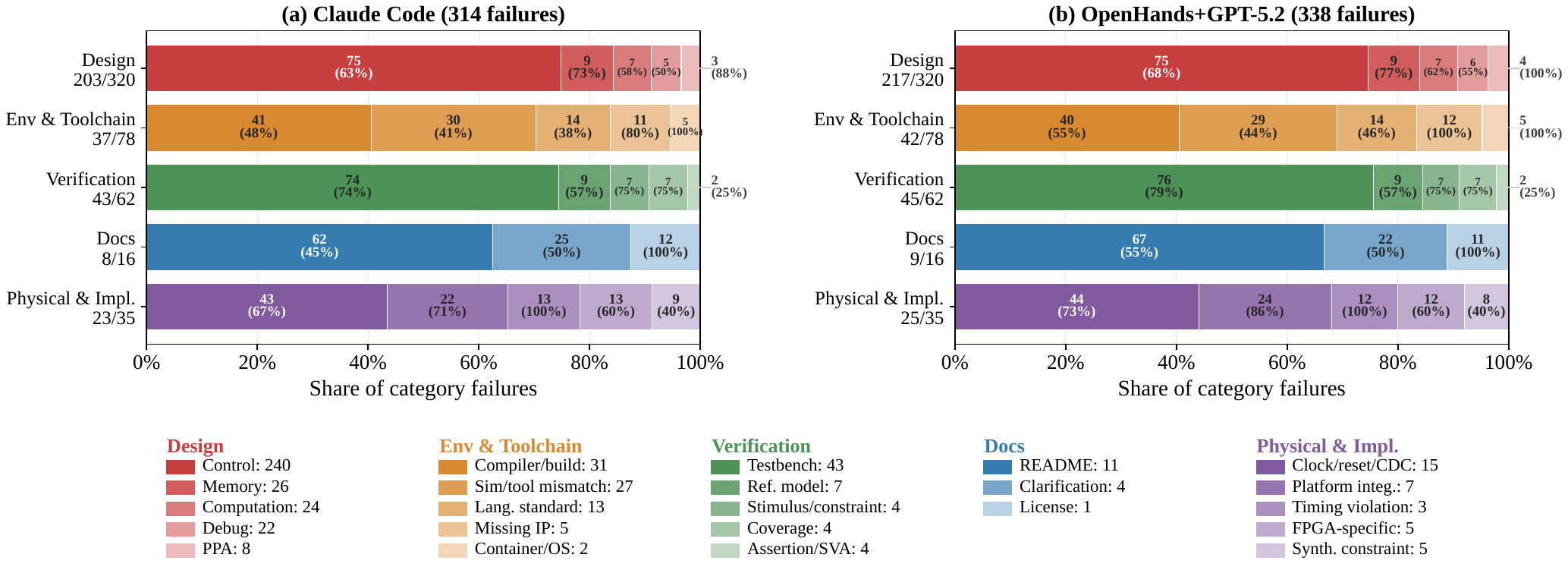}
  \caption{Per-category failure breakdown into fine-grained subcategories, for (a) Claude Code and (b) OpenHands+GPT-5.2. Row labels give failed/total instances in each main category. For each segment, the upper number is the subcategory's share of that category's failures (\%); the parenthesized number is its failure rate. Legend counts give the number of instances in each subcategory across the full 511-instance corpus. PPA: power, performance, and area; CDC: clock-domain crossing; OS: operating system.}
  \label{fig:failure_subcategory}
\end{figure*}

Figure~\ref{fig:failure_category} shows two breakdowns of failures on the same 511-case corpus: by issue category in panel (a), and by failure stage in panel (b). Each failure is tagged \emph{No Edit} (0 HDL files touched), \emph{Localization Failure} (HDL files edited but not all gold-patch files covered), or \emph{Repair Failure} (every gold file reached but the fix is wrong). Both Claude Code and OpenHands+GPT-5.2 land in the same $77\%/23\%$ upstream/Repair split ($29.9\%$ No Edit + $47.1\%/47.3\%$ Localization vs.\ $22.9\%/22.8\%$ Repair); design issues account for $203/314$ and $217/338$ failures ($64.6\%/64.2\%$), followed by verification ($43/314$ and $45/338$, or $13.7\%/13.3\%$). Figure~\ref{fig:failure_subcategory} drills into each issue category and shows that the failure mass is concentrated in a small number of subcategories. Control-flow / FSM bugs produce $152/203$ Claude and $162/217$ OpenHands design failures (within-subcategory failure rates $63\%/68\%$); testbench bugs produce $32/43$ and $34/45$ verification failures ($74\%/79\%$); Env-and-Toolchain failures concentrate in compiler/build and simulator-mismatch issues, while Physical-and-Implementation is dominated by clock/reset/clock-domain crossing (CDC) and platform-integration. Per-segment within-subcategory failure rates remain in the $25$ to $100$\% band on both panels.

\finding{Agents fail mostly on design control-flow / FSM bugs and on verification testbench bugs, with physical failures clustered on clock/reset/CDC and platform integration; env-and-toolchain failures are a long tail of compiler/build, simulator-mismatch, and language-standard issues.}

\subsection{How does performance vary with patch complexity?}
\label{sec:complexity}

We partition the 511 instances by the HDL files and raw unified-diff hunks in their gold patches: \emph{T1 Easy} (one HDL file with exactly one hunk), \emph{T2 Medium} (one HDL file with 2 to 5 hunks), and \emph{T3 Hard} (at least 6 HDL hunks in total or edits to more than one HDL file). In this classification, \emph{cross-hierarchy} denotes edits to HDL files in more than one top-level directory; any multi-file HDL patch belongs to T3, including edits within the same directory. The exact rules and a coarser file-and-line-based difficulty distribution are in Appendix~\ref{app:patch-complexity}. Figure~\ref{fig:diag_tier_oracle}(a) shows that OpenAI Codex, Claude Code, and OpenHands+GPT-5.2 all drop monotonically by roughly $1.8\times$ from T1 to T3 (Codex $53.9 \to 30.5$, Claude $53.9 \to 29.8$, OpenHands $47.7 \to 26.7$); the T3 absolute level ($\approx 27$ to $31$\%) sits more than $20$\% below each agent's own T1. The inter-agent gap is $<8\%$ at every tier, Codex and Claude tie at T1, Claude leads at T2 ($45.1\%$ vs.\ $39.6\%$), and Codex edges Claude back at T3 ($30.5\%$ vs.\ $29.8\%$); Claude's overall lead ($38.6\%$ vs.\ Codex $38.0\%$ vs.\ OpenHands $33.9\%$) is driven by T2.
\finding{All three agents resolve fewer than one third of T3 cases, whose gold patches contain at least six HDL hunks or edits to multiple HDL files.}

\begin{figure}[t]
  \centering
  \begin{minipage}[b]{0.42\linewidth}
    \centering
    \includegraphics[width=\linewidth]{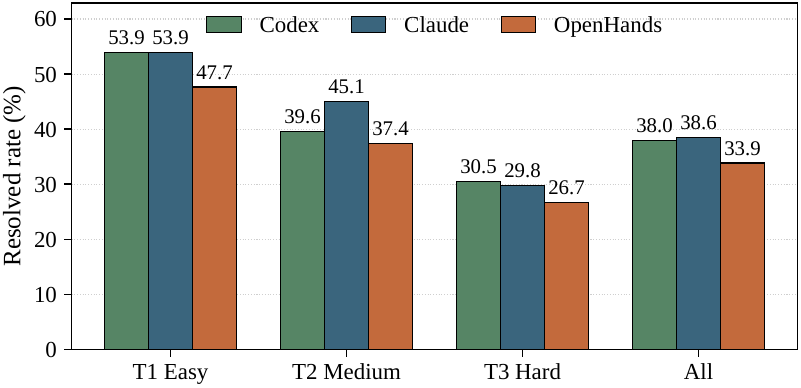}\\[2pt]
    {\small (a)}
  \end{minipage}\hfill
  \begin{minipage}[b]{0.30\linewidth}
    \centering
    \includegraphics[width=\linewidth]{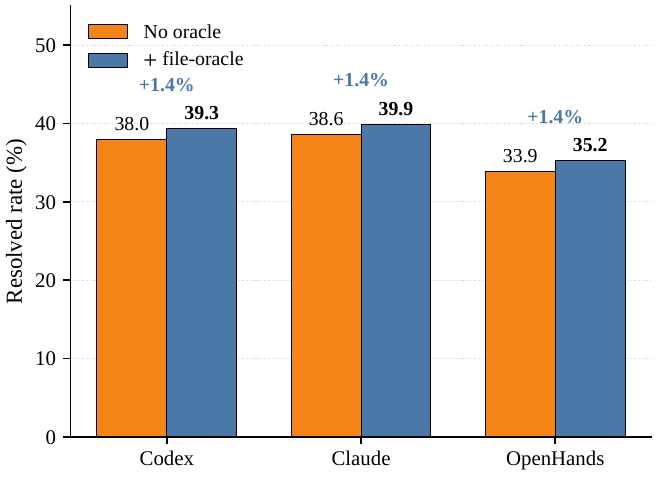}\\[2pt]
    {\small (b)}
  \end{minipage}\hfill
  \begin{minipage}[b]{0.26\linewidth}
    \centering
    \includegraphics[width=\linewidth]{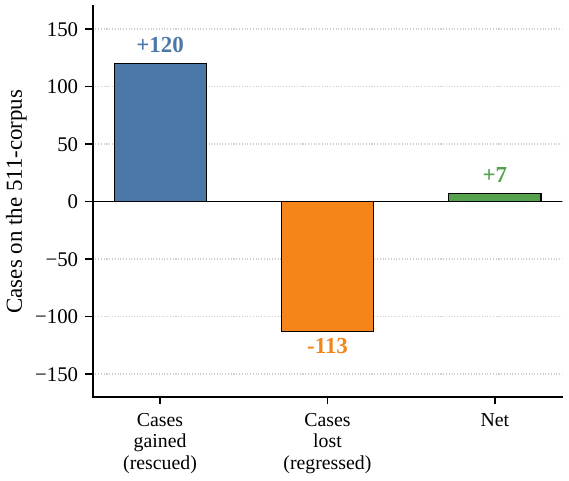}\\[2pt]
    {\small (c)}
  \end{minipage}
  \caption{(a) Resolved rate by patch-complexity tier (b) Resolved rate without and with file-level oracle. (c) Per-case decomposition on Claude: rescues vs.\ regressions.}
  \label{fig:diag_tier_oracle}
\end{figure}

\subsection{How much can better file localization help?}
\label{sec:oracle}

We give each agent the gold-patch file list at prompt time and otherwise leave the editing pipeline unchanged. Figure~\ref{fig:diag_tier_oracle}(b) shows that Claude Code moves only from $38.6\%$ to $39.9\%$ ($\Delta = +1.4$ percentage points); Codex and OpenHands+GPT-5.2 land within $\pm 0.1$ percentage points of the same $\Delta$, consistent with the gain and regression mechanisms (panel (c)) being largely backbone-agnostic. Figure~\ref{fig:diag_tier_oracle}(c) decomposes Claude's net $+7$ cases into $+120$ rescued failures (the file hint genuinely unblocks the agent) and $-113$ regressions on baseline-PASS snapshots; four mechanisms explain the offset: a file hint says nothing about which line, operator, or direction to change so the agent's wrong-guess rate consumes most of the gain; the file list reads as ``edit something,'' and the oracle run turns $113$ of the $197$ baseline-successful cases into failures ($113/197 = 57.4\%$); a flat list omits the cross-module propagation chain, so the agent fixes the listed file but leaves upstream/downstream files alone; and the oracle cannot express ``this file is correctly silent,'' so a baseline-PASS check before editing would avoid most regressions.
\finding{Telling the agent which file to edit barely helps, because the agent then breaks files that did not need editing; useful localization hints must be finer than the file level.}

\subsection{How much does testbench feedback contribute?}
\label{sec:eda_feedback}

Unlike a flat file list, Phoenix-bench's verifier emits Verilator compile and runtime logs from the buggy phase that go strictly beyond ``which file to edit'': they additionally pinpoint the offending line and operator, name the implicated module along its instantiation chain, and supply the expected widths/types/values that the bug violates. We contrast (i) \emph{no feedback}, where the agent receives only issue text and repository, with (ii) \emph{$+$ testbench feedback}, where the same agent reads Verilator's log after each attempt and applies at most one corrective re-edit. Table~\ref{tab:eda_feedback}(a) shows that all three agents gain $42.1$--$45.4$ percentage points, with Claude Code moving from $197/511$ to $429/511$ ($38.6\% \to 84.0\%$, $+45.4$ percentage points), Codex from $194/511$ to $419/511$ ($+44.0$ percentage points), and OpenHands+GPT-5.2 from $173/511$ to $388/511$ ($+42.1$ percentage points). Table~\ref{tab:eda_feedback}(b) shows that on Claude's 314 failures the rescue rate is monotone across the three pre-feedback failure stages: $77.7\%$ for \emph{No Edit} ($73/94$), $73.6\%$ for \emph{Localization Failures} ($109/148$), and $69.4\%$ for \emph{Repair Failures} ($50/72$). These stages account for all $73+109+50=232$ additional resolved cases. The pattern holds across design, verification, physical-implementation, and docs categories with one exception: env-and-toolchain failures see only $\sim 60\%$ rescue, because the underlying bug is in tool or simulator semantics and Verilator's log can be misleading or empty.
\finding{Letting the agent read the verifier's error log helps far more than telling it which file to edit, because the log says \emph{where} the bug is and \emph{what} the fix has to look like.}

\begin{table}[t]
\centering
\caption{Test feedback impact. \textbf{(a)} Resolved rate before  vs.\ after one round of feedback for the three interactive agents; $\Delta$ is in percentage points. \textbf{(b)} Rescue rate by Claude's pre-feedback failure stage on the 314 failures.}
\label{tab:eda_feedback}
\footnotesize
\setlength{\tabcolsep}{3pt}
\begin{minipage}[t]{0.58\linewidth}
\centering
\textbf{(a) Per-agent resolved rate}\\[2pt]
\begin{tabular}{lccc}
\toprule
\textbf{Agent} & \textbf{No fb.} & \textbf{$+$ TB fb.} & \textbf{$\Delta$} \\
\midrule
OpenAI Codex      & 194/511 (38.0\%) & 419/511 (82.0\%) & $+44.0$ \\
Claude Code       & 197/511 (38.6\%) & 429/511 (84.0\%) & $+45.4$ \\
OpenHands+GPT-5.2 & 173/511 (33.9\%) & 388/511 (75.9\%) & $+42.1$ \\
\bottomrule
\end{tabular}
\end{minipage}\hfill
\begin{minipage}[t]{0.40\linewidth}
\centering
\textbf{(b) Claude rescue rate by stage}\\[2pt]
\begin{tabular}{lcc}
\toprule
\textbf{Pre-feedback stage} & \textbf{Rescued} & \textbf{Rate} \\
\midrule
No Edit               & 73/94   & 77.7\% \\
Localization Failure  & 109/148 & 73.6\% \\
Repair Failure        & 50/72   & 69.4\% \\
\bottomrule
\end{tabular}
\end{minipage}
\end{table}

\section{Conclusion}

We introduced \textbf{Phoenix-bench}, an execution-grounded hardware-issue-resolution benchmark of 511 verified Verilator instances from 114 GitHub repositories, and ran a uniform evaluation of four deployed product agents and eight open-source agentic structures across four LLM backbones plus two diagnostic interventions. (i)~Software and hardware are fundamentally different engineering tasks: the same agent loses $37$ to $58$ percentage points from SWE-bench Verified to Phoenix-bench because hardware bugs propagate across parallel instantiated modules through signal flow, not along a software-style call graph. (ii)~Failures concentrate on design control-flow / FSM bugs, verification testbench bugs, and T3 cases whose gold patches contain at least six HDL hunks or edits to multiple HDL files. (iii)~Localization granularity, not localization itself, is the controllable lever: a perfect file-level oracle adds only $+1.4$\% because the agent then breaks files it should not edit, whereas one round of Verilator-log feedback adds $42.1$--$45.4$ percentage points because the log says \emph{where} the bug is and \emph{what} the fix has to look like.

\begin{ack}
This research is supported by the Ministry of Education, Singapore, under its Academic Research Fund Tier 3 grant MOE-MOET32024-0003 and Tier 2 grant T2EP20224-0038. Any opinions, findings, and conclusions or recommendations expressed in this material are those of the authors and do not reflect the views of the Ministry of Education, Singapore. We also thank AMD for the generous donation of the Heterogeneous Accelerated Compute Cluster (HACC) at the National University of Singapore. Apart from the support disclosed above, the authors declare no competing interests.
\end{ack}

\bibliographystyle{plainnat}
\bibliography{reference}

@String{Computing = "Computing" }

@String{Computer = "{IEEE} Computer" }

@article{thakur2024verigen,
  title = {{VeriGen: A Large Language Model for Verilog Code Generation}},
  author = {Thakur, Shailja and Ahmad, Baleegh and Pearce, Hammond and Tan, Benjamin and Dolan-Gavitt, Brendan and Karri, Ramesh and Garg, Siddharth},
  journal = {ACM Transactions on Design Automation of Electronic Systems},
  year = {2024},
  volume = {29},
  number = {3},
  publisher = {Association for Computing Machinery (ACM)},
  doi = {10.1145/3643681},
  url = {https://doi.org/10.1145/3643681},
  articleno = {46},
  numpages = {31},
}

@inproceedings{liu2023verilogeval,
  title = {{Invited Paper: VerilogEval: Evaluating Large Language Models for Verilog Code Generation}},
  author = {Liu, Mingjie and Pinckney, Nathaniel and Khailany, Brucek and Ren, Haoxing},
  booktitle = {2023 IEEE/ACM International Conference on Computer Aided Design},
  year = {2023},
  pages = {1--8},
  publisher = {IEEE},
  doi = {10.1109/iccad57390.2023.10323812},
  url = {https://doi.org/10.1109/iccad57390.2023.10323812},
}

@inproceedings{yao2025location,
  title = {{Location is Key: Leveraging LLM for Functional Bug Localization in Verilog Design}},
  author = {Yao, Bingkun and Wang, Ning and Zhou, Jie and Wang, Xi and Gao, Hong and Jiang, Zhe and Guan, Nan},
  booktitle = {2025 62nd ACM/IEEE Design Automation Conference},
  year = {2025},
  pages = {1--7},
  publisher = {IEEE},
  doi = {10.1109/dac63849.2025.11133280},
  url = {https://doi.org/10.1109/dac63849.2025.11133280},
}

@inproceedings{zhou2025insights,
  title = {{Insights from Rights and Wrongs: A Large Language Model for Solving Assertion Failures in RTL Design}},
  author = {Zhou, Jie and Ji, Youshu and Wang, Ning and Hu, Yuchen and Jiao, Xinyao and Yao, Bingkun and Fang, Xinwei and Zhao, Shuai and Guan, Nan and Jiang, Zhe},
  booktitle = {2025 62nd ACM/IEEE Design Automation Conference},
  year = {2025},
  pages = {1--7},
  publisher = {IEEE},
  doi = {10.1109/dac63849.2025.11133324},
  url = {https://doi.org/10.1109/dac63849.2025.11133324},
}

@inproceedings{fu2025hwfixbench,
  title = {{HWFixBench: Benchmarking Tools for Hardware Understanding and Fault Repair}},
  author = {Fu, Weimin and Li, Shijie and Jin, Yier and Guo, Xiaolong},
  booktitle = {Proceedings of the Great Lakes Symposium on VLSI 2025},
  year = {2025},
  pages = {427--434},
  publisher = {ACM},
  doi = {10.1145/3716368.3735171},
  url = {https://doi.org/10.1145/3716368.3735171},
}

@inproceedings{jimenez2023swe,
  title = {{SWE-bench: Can Language Models Resolve Real-World GitHub Issues?}},
  author = {Jimenez, Carlos E. and Yang, John and Wettig, Alexander and Yao, Shunyu and Pei, Kexin and Press, Ofir and Narasimhan, Karthik R.},
  booktitle = {The Twelfth International Conference on Learning Representations},
  year = {2024},
  url = {https://openreview.net/forum?id=VTF8yNQM66},
}

@inproceedings{lu2024rtllm,
  title = {{RTLLM: An Open-Source Benchmark for Design RTL Generation with Large Language Model}},
  author = {Lu, Yao and Liu, Shang and Zhang, Qijun and Xie, Zhiyao},
  booktitle = {2024 29th Asia and South Pacific Design Automation Conference},
  year = {2024},
  pages = {722--727},
  publisher = {IEEE},
  doi = {10.1109/asp-dac58780.2024.10473904},
  url = {https://doi.org/10.1109/asp-dac58780.2024.10473904},
}

@inproceedings{batten2024pyhdl,
  title = {{PyHDL-Eval: An LLM Evaluation Framework for Hardware Design Using Python-Embedded DSLs}},
  author = {Batten, Christopher and Pinckney, Nathaniel and Liu, Mingjie and Ren, Haoxing and Khailany, Brucek},
  booktitle = {Proceedings of the 2024 ACM/IEEE International Symposium on Machine Learning for CAD},
  year = {2024},
  pages = {1--17},
  publisher = {ACM},
  doi = {10.1145/3670474.3685948},
  url = {https://doi.org/10.1145/3670474.3685948},
}

@inproceedings{kashanaki2024hdleval,
  title = {{HDLEval Benchmarking LLMs for multiple HDLs}},
  author = {Kashanaki, Farzaneh Rabiei and Zakharov, Mark and Renau, Jose},
  booktitle = {2024 IEEE LLM Aided Design Workshop},
  year = {2024},
  pages = {1--5},
  publisher = {IEEE},
  doi = {10.1109/lad62341.2024.10691770},
  url = {https://doi.org/10.1109/lad62341.2024.10691770},
}

@inproceedings{yang2024swe,
  title = {{SWE-agent: Agent-Computer Interfaces Enable Automated Software Engineering}},
  author = {Yang, John and Jimenez, Carlos E. and Wettig, Alexander and Lieret, Kilian and Yao, Shunyu and Narasimhan, Karthik and Press, Ofir},
  booktitle = {Advances in Neural Information Processing Systems},
  volume = {37},
  pages = {50528--50652},
  year = {2024},
  publisher = {Curran Associates, Inc.},
  doi = {10.52202/079017-1601},
  url = {https://proceedings.neurips.cc/paper_files/paper/2024/hash/5a7c947568c1b1328ccc5230172e1e7c-Abstract-Conference.html},
}

@inproceedings{wang2024openhands,
  title = {{OpenHands: An Open Platform for AI Software Developers as Generalist Agents}},
  author = {Wang, Xingyao and Li, Boxuan and Song, Yufan and Xu, Frank F. and Tang, Xiangru and Zhuge, Mingchen and Pan, Jiayi and Song, Yueqi and Li, Bowen and Singh, Jaskirat and Tran, Hoang and Li, Fuqiang and Ma, Ren and Zheng, Mingzhang and Qian, Bill and Shao, Daniel and Muennighoff, Niklas and Zhang, Yizhe and Hui, Binyuan and Lin, Junyang and Brennan, Robert and Peng, Hao and Ji, Heng and Neubig, Graham},
  booktitle = {The Thirteenth International Conference on Learning Representations},
  year = {2025},
  url = {https://proceedings.iclr.cc/paper_files/paper/2025/hash/a4b6ad6b48850c0c331d1259fc66a69c-Abstract-Conference.html},
}

@article{xia2024agentless,
  title = {{Demystifying LLM-Based Software Engineering Agents}},
  author = {Xia, Chunqiu Steven and Deng, Yinlin and Dunn, Soren and Zhang, Lingming},
  journal = {Proceedings of the ACM on Software Engineering},
  year = {2025},
  volume = {2},
  number = {FSE},
  pages = {801--824},
  publisher = {Association for Computing Machinery (ACM)},
  doi = {10.1145/3715754},
  url = {https://doi.org/10.1145/3715754},
}

@misc{openai_5_2_system_card,
  title = {{Update to GPT-5 System Card: GPT-5.2}},
  author = {{OpenAI}},
  year = {2025},
  url = {https://cdn.openai.com/pdf/3a4153c8-c748-4b71-8e31-aecbde944f8d/oai_5_2_system-card.pdf},
  note = {Accessed 2026-10-30},
}

@misc{google_gemini_3_pro,
  title = {{Gemini 3 Pro Model Card}},
  author = {{Google DeepMind}},
  year = {2026},
  url = {https://deepmind.google/models/model-cards/gemini-3-pro/},
  note = {Model released November 2025; model card updated May 2026. Accessed 2026-10-30},
}

@misc{liu2025deepseek,
  title = {{DeepSeek-V3.2: Pushing the Frontier of Open Large Language Models}},
  author = {{DeepSeek-AI} and Liu, Aixin and Mei, Aoxue and Lin, Bangcai and others},
  year = {2025},
  eprint = {2512.02556},
  archivePrefix = {arXiv},
  doi = {10.48550/arXiv.2512.02556},
  url = {https://arxiv.org/abs/2512.02556},
}

@misc{zou2026mhrc,
  title = {{MHRC-Bench: A Multilingual Hardware Repository-Level Code Completion Benchmark}},
  author = {Zou, Qingyun and Cui, Jiahao and Chen, Nuo and He, Bingsheng and Wong, Weng-Fai},
  year = {2026},
  eprint = {2601.03708},
  archivePrefix = {arXiv},
  doi = {10.48550/arXiv.2601.03708},
  url = {https://arxiv.org/abs/2601.03708},
}

@inproceedings{tsai2024rtlfixer,
  title = {{RTLFixer: Automatically Fixing RTL Syntax Errors with Large Language Model}},
  author = {Tsai, Yunda and Liu, Mingjie and Ren, Haoxing},
  booktitle = {Proceedings of the 61st ACM/IEEE Design Automation Conference},
  year = {2024},
  pages = {1--6},
  publisher = {ACM},
  doi = {10.1145/3649329.3657353},
  url = {https://doi.org/10.1145/3649329.3657353},
}

@article{mu2025experepair,
  title = {{ExpeRepair: Dual-Memory Enhanced LLM-Based Repository-Level Program Repair}},
  author = {Mu, Fangwen and Wang, Junjie and Shi, Lin and Wang, Song and Li, Shoubin and Wang, Qing},
  journal = {Proceedings of the ACM on Software Engineering},
  year = {2026},
  volume = {3},
  number = {FSE},
  pages = {3954--3976},
  publisher = {Association for Computing Machinery (ACM)},
  doi = {10.1145/3808181},
  url = {https://doi.org/10.1145/3808181},
}

@misc{miniswe2025,
  title = {{mini-SWE-agent: The 100-Line AI Agent That Solves GitHub Issues}},
  author = {{SWE-agent Team}},
  year = {2025},
  url = {https://github.com/SWE-agent/mini-swe-agent},
  note = {Accessed 2026-10-30},
}

@misc{kgcompass2025,
  title = {{Enhancing Repository-Level Software Repair via Repository-Aware Knowledge Graphs}},
  author = {Yang, Boyang and Ren, Jiadong and Jin, Shunfu and Liu, Yang and Liu, Feng and Le, Bach and Tian, Haoye},
  year = {2025},
  eprint = {2503.21710},
  archivePrefix = {arXiv},
  doi = {10.48550/arXiv.2503.21710},
  url = {https://arxiv.org/abs/2503.21710},
}

@misc{lingxi2025,
  title = {{Lingxi: Knowledge-Guided Multi-Agent Framework for Repository-Level Issue Resolution}},
  author = {{Lingxi Team}},
  year = {2025},
  url = {https://github.com/lingxi-agent/Lingxi},
  note = {Accessed 2026-10-30},
}

@misc{aider2024,
  title = {{Aider: AI Pair Programming in Your Terminal}},
  author = {Gauthier, Paul},
  year = {2024},
  url = {https://aider.chat/},
  note = {Accessed 2026-10-30},
}

@article{chen2024hdldebugger,
  title = {{HDLdebugger: Streamlining HDL debugging with Large Language Models}},
  author = {Yao, Xufeng and Li, Haoyang and Chan, Tsz Ho and Xiao, Wenyi and Yuan, Mingxuan and Huang, Yu and Chen, Lei and Yu, Bei},
  journal = {ACM Transactions on Design Automation of Electronic Systems},
  year = {2025},
  volume = {30},
  number = {6},
  publisher = {Association for Computing Machinery (ACM)},
  doi = {10.1145/3735638},
  url = {https://doi.org/10.1145/3735638},
  articleno = {102},
  numpages = {26},
}

@article{ahmad2023fixing,
  title = {{On Hardware Security Bug Code Fixes by Prompting Large Language Models}},
  author = {Ahmad, Baleegh and Thakur, Shailja and Tan, Benjamin and Karri, Ramesh and Pearce, Hammond},
  journal = {IEEE Transactions on Information Forensics and Security},
  year = {2024},
  volume = {19},
  pages = {4043--4057},
  publisher = {Institute of Electrical and Electronics Engineers (IEEE)},
  doi = {10.1109/tifs.2024.3374558},
  url = {https://doi.org/10.1109/tifs.2024.3374558},
}

@inproceedings{fang2024assertllm,
  title = {{AssertLLM: Generating Hardware Verification Assertions from Design Specifications via Multi-LLMs}},
  author = {Yan, Zhiyuan and Fang, Wenji and Li, Mengming and Li, Min and Liu, Shang and Xie, Zhiyao and Zhang, Hongce},
  booktitle = {Proceedings of the 30th Asia and South Pacific Design Automation Conference},
  year = {2025},
  pages = {614--621},
  publisher = {ACM},
  doi = {10.1145/3658617.3697756},
  url = {https://doi.org/10.1145/3658617.3697756},
}

@misc{anthropic_claude_code_2026,
  title = {{Claude Code Overview}},
  author = {{Anthropic}},
  year = {2026},
  url = {https://code.claude.com/docs/en/overview},
  note = {Accessed 2026-10-30},
}

@misc{openai_codex_2026,
  title = {{Codex}},
  author = {{OpenAI}},
  year = {2026},
  url = {https://openai.com/codex/},
  note = {Accessed 2026-10-30},
}

@misc{github_copilot_coding_agent_2026,
  title = {{GitHub Copilot on GitHub.com and GitHub Mobile}},
  author = {{GitHub}},
  year = {2026},
  url = {https://docs.github.com/en/copilot/concepts/copilot-surfaces/copilot-on-github},
  note = {Accessed 2026-10-30},
}

@misc{google_gemini_cli_2026,
  title = {{Gemini CLI}},
  author = {{Google}},
  year = {2026},
  url = {https://github.com/google-gemini/gemini-cli},
  note = {Accessed 2026-10-30},
}

@misc{openai_swebench_verified_2024,
  title = {{Introducing SWE-bench Verified}},
  author = {Chowdhury, Neil and Aung, James and Shern, Chan Jun and Jaffe, Oliver and Sherburn, Dane and Starace, Giulio and Mays, Evan and Dias, Rachel and Aljubeh, Marwan and Glaese, Mia and Jimenez, Carlos E. and Yang, John and Ho, Leyton and Patwardhan, Tejal and Liu, Kevin and Madry, Aleksander},
  year = {2024},
  url = {https://openai.com/index/introducing-swe-bench-verified/},
  note = {Accessed 2026-10-30},
}

@misc{zou2026hlsseek,
  title = {{HLS-Seek: QoR-Aware Code Generation for High-Level Synthesis via Proxy Comparative Reward Reinforcement Learning}},
  author = {Zou, Qingyun and Yu, Feng and Tan, Hongshi and Chen, Yao and He, Bingsheng and Wong, Weng-Fai},
  year = {2026},
  eprint = {2605.13536},
  archivePrefix = {arXiv},
  doi = {10.48550/arXiv.2605.13536},
  url = {https://arxiv.org/abs/2605.13536},
  note = {Accepted to the 2026 45th IEEE/ACM International Conference on Computer-Aided Design. Program: \url{https://iccad.com/2026/full-program}},
}

@misc{zou2026rwopd,
  title = {{Reward-Weighted On-Policy Distillation with an Open Property-Equivalence Verifier for NL-to-SVA Generation}},
  author = {Zou, Qingyun and Li, Yingze and Liu, Tianen and He, Bingsheng and Wong, Weng-Fai},
  year = {2026},
  eprint = {2605.13501},
  archivePrefix = {arXiv},
  doi = {10.48550/arXiv.2605.13501},
  url = {https://arxiv.org/abs/2605.13501},
}

@inproceedings{allam2024rtlrepo,
  title = {{RTL-Repo: A Benchmark for Evaluating LLMs on Large-Scale RTL Design Projects}},
  author = {Allam, Ahmed and Shalan, Mohamed},
  booktitle = {2024 IEEE LLM Aided Design Workshop},
  year = {2024},
  pages = {1--5},
  publisher = {IEEE},
  doi = {10.1109/lad62341.2024.10691810},
  url = {https://doi.org/10.1109/lad62341.2024.10691810},
}

@misc{qwen3coder2025,
  title = {{Qwen3-Coder: Agentic Coding in the World}},
  author = {{Qwen Team}},
  year = {2025},
  url = {https://qwenlm.github.io/blog/qwen3-coder/},
  note = {Accessed 2026-10-30},
}

@misc{swebench_leaderboard_2026,
  title = {{SWE-bench Leaderboards}},
  author = {{SWE-bench Team}},
  year = {2026},
  url = {https://www.swebench.com/},
  note = {Accessed 2026-10-30},
}

@misc{shi2026earl,
  title = {{EARL: Entropy-Aware RL Alignment of LLMs for Reliable RTL Code Generation}},
  author = {Shi, Jiahe and Gao, Zhengqi and Ko, Ching-Yun and Boning, Duane},
  year = {2026},
  eprint = {2511.12033},
  archivePrefix = {arXiv},
  primaryClass = {cs.LG},
  doi = {10.48550/arXiv.2511.12033},
  url = {https://arxiv.org/abs/2511.12033v2},
  note = {Accepted to the 2026 63rd ACM/IEEE Design Automation Conference. Program entry: \url{https://63dac.conference-program.com/presentation/?id=RESEARCH1612&sess=sess314}},
}

@misc{zou2025hlstrans,
  title = {{HLStrans: Dataset for C-to-HLS Hardware Code Synthesis}},
  author = {Zou, Qingyun and Chen, Nuo and Chen, Yao and He, Bingsheng and Wong, Weng-Fai},
  year = {2025},
  eprint = {2507.04315},
  archivePrefix = {arXiv},
  doi = {10.48550/arXiv.2507.04315},
  url = {https://arxiv.org/abs/2507.04315},
}

\appendix
\clearpage

\section{Patch Complexity Classification Rules}
\label{app:patch-complexity}

\paragraph{Coarse difficulty distribution.}
For the 511 released instances, a coarser stratification counts distinct edited file paths across all file types and sums added and deleted lines within patch hunks. We assign each patch to one of three disjoint bands: \emph{Easy} (one file and at most 10 edited lines, 97 cases averaging 1.0 files and 4.1 edited lines), \emph{Medium} (at most three files and at most 50 edited lines, excluding Easy cases; 109 cases averaging 1.8 files and 19.5 edited lines), and \emph{Hard} (at least four files or more than 50 edited lines, 305 cases). The Hard split is heavy-tailed: it averages 24.4 files and 3{,}349.4 edited lines, with a median of 205 edited lines because some repository updates include very large generated or vendored diffs. Overall, $414/511$ ($81.0\%$) of instances fall in Medium or Hard, requiring multi-file or substantial edits.

To stratify the 511-case Phoenix-bench corpus by developer-patch structure
(\S\ref{sec:complexity}, tiers T1/T2/T3), we apply the following rule set to each
gold patch. The rules operate purely on the unified diff and consider
only HDL source files; scripts, Makefiles, testbenches, and documentation
are ignored.

\paragraph{Inputs considered.}
\begin{itemize}
    \item \textbf{File filter.} Only files whose extension is one of
    \texttt{.v}, \texttt{.sv}, \texttt{.vh}, or \texttt{.svh} are counted.
    All other paths in the diff are skipped.
    \item \textbf{Hunk count.} Each unified-diff hunk header
    (\texttt{@@~-...~+...~@@}) within an HDL file contributes one to the
    hunk count. Hunks are not merged across line gaps; this matches the
    raw \texttt{git diff} segmentation.
    \item \textbf{File count.} The number of distinct HDL file paths
    appearing under \texttt{diff~--git} entries.
    \item \textbf{Hierarchy.} The \emph{top-level directory} of each HDL
    path (\texttt{Path.parts[0]}). A patch is \emph{cross-hierarchy} when
    it touches files in more than one top-level directory.
\end{itemize}

\paragraph{Tier rules (evaluated in order).}
\begin{enumerate}
    \item If total HDL hunks $\geq 6$: \textbf{T3 Hard}.
    \item Else if HDL files $> 1$ \emph{and} cross-hierarchy:
    \textbf{T3 Hard}.
    \item Else if HDL files $> 1$ (same top-level directory):
    \textbf{T3 Hard} (multi-file).
    \item Else if a single HDL file with exactly $1$ hunk:
    \textbf{T1 Easy}.
    \item Else if a single HDL file with $2$ to $5$ hunks:
    \textbf{T2 Medium}.
    \item Otherwise: T3 Hard (fallback; not exercised in the released
    corpus).
\end{enumerate}

\paragraph{Resulting distribution.}
Applying these rules to the 511 verified instances yields
\textbf{T1: 128 (25.0\%)}, \textbf{T2: 91 (17.8\%)}, and
\textbf{T3: 292 (57.1\%)}. The two patches in the curated set without a
matching Verilator snapshot
(\texttt{FPGA-MAFIA\_fpga\_mafia\_pr415}, \texttt{UTOSS\_risc-v\_pr35})
are excluded.

\paragraph{Notes and caveats.}
The rules deliberately use the raw \texttt{@@} hunk segmentation rather
than a semantic block merge: a developer change that spans a declaration
and its later use will appear as two hunks even when it is a single
logical edit, which inflates T2/T3 relative to a "logical block" count.
Likewise, any patch that touches more than one HDL file is promoted to
T3, even when the second file is a tightly coupled header (\texttt{.svh})
of the same module. These choices keep the classifier deterministic and
auditable from the diff alone, but they make the released distribution
skew more conservative (more T3) than a semantically merged count would.

\section{Limitations}
\label{app:limitations}

\paragraph{Task text.}
Issue text varies in detail. Of the 511 tasks, 37 contain only a short title and 474 contain more
issue text. More text does not always mean that every repair detail is present: an independent
judge marked 26.5\% of a 102-task sample as missing key details. We therefore report this check
separately and avoid treating every task as equally specified.

\paragraph{Tool and design-flow scope.}
The current package uses repeatable open-source EDA checks, mainly Verilator. It does not cover
every SystemVerilog feature, very high-speed integrated circuit hardware description language (VHDL), proprietary simulators, or full place-and-route and timing
sign-off flows. Results may change when these tools and stages are added.

\paragraph{Models and runs.}
Product agents are tested at fixed versions and dates. The reported results describe these runs
rather than all future versions of the products. Repeated runs are also needed to measure
run-to-run variation.

\paragraph{Patterns and causes.}
Failure stages and case studies show where agents stop, but they do not prove why. Signal-flow
search and software-focused training are possible reasons, not identified causes. The feedback
experiment also gives several kinds of information at once, so it does not measure the value of
each part separately.

\paragraph{Test gaming audit.}
For product-agent runs, the checker and run scripts are outside the editable workspace. We also
scan recovered patches for checker edits, hard-coded checked values, removed assertions, and empty
patches counted as passes. Across 749 successful product-agent patches, we found no confirmed
scored pass caused by these signs; the one-sided 95\% upper bound is 0.40\%. An empty patch passes
0/511 tasks, and the developer patch passes 511/511. This audit does not rule out every form of
gaming.

\section{Case study: \texttt{hpdcache\_pr33}}
\label{app:case_study}

\begin{figure}[t]
  \centering
  \includegraphics[width=\linewidth]{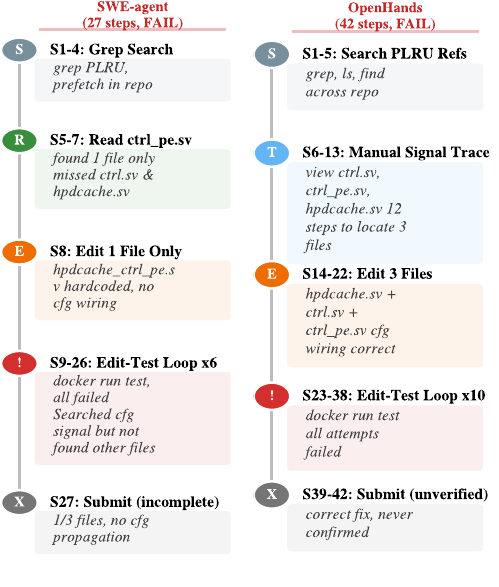}
  \caption{Case study on \texttt{hpdcache\_pr33}: a cross-module signal-propagation issue that requires wiring \texttt{cfg\_prefetch\_updt\_plru} through the hierarchy. PLRU: pseudo-least-recently-used.}
  \label{fig:case_study}
\end{figure}

Figure~\ref{fig:case_study} illustrates a Phoenix-bench case from HPDcache that concretizes the hierarchy / parameter failure category discussed in \S\ref{sec:hdl_taxonomy}. The issue requires wiring a configuration signal through three files in the module hierarchy: \texttt{hpdcache.sv} to \texttt{ctrl.sv} to \texttt{ctrl\_pe.sv}. A correct repair extracts the signal from the issue, traces the propagation chain across modules, and applies one coordinated edit. Keyword search alone can find the downstream consumer while missing upstream modules where the new signal is not yet mentioned. Even oracle file localization (\S\ref{sec:oracle}) is insufficient here because the agent must still construct the port chain rather than merely identify the file set. This case demonstrates why realistic hardware issue resolution requires understanding signal-flow dependencies, not only finding lexical matches.

\section{Artifact, Licensing, and Reproducibility Package}
\label{app:artifact}

The code repository is available at
\url{https://github.com/zanehpn/Phoenix-Bench}, with benchmark data at
\url{https://huggingface.co/datasets/OvertureMMXXI/Phoenix-Bench}.
The data repository holds the 511 benchmark instances, and the code repository
holds the agent frameworks and the verifier scripts. The records below describe
what each release contains.

\paragraph{Instance index and licensing.}
The file \texttt{data/test.jsonl} in the data repository indexes the 511
instances, and the accompanying Croissant metadata lists each instance's
identifier with its issue, developer-patch, and testbench links. Each identifier
names the upstream owner, repository, and pull request. Phoenix-bench's own
contribution is released under the MIT license, while each redistributed source
snapshot remains governed by its upstream license.

\paragraph{Artifact layout and evaluation items.}
The released package mirrors the internal benchmark layout used for the
experiments: \texttt{issues/} stores the issue metadata and prompt-facing
problem statement, \texttt{snapshots/} stores the pinned repository source
snapshots, \texttt{patches/} stores the developer patches, and
\texttt{testbench/} stores the executable runners and Verilator
testbenches. Each evaluation item links one issue record, one source snapshot,
one developer patch, and one runner directory. The runner directory normally
contains two audit logs: a \emph{buggy log}, which records the compile or
runtime error observed before the fix, and a \emph{pass log}, which records the
same verifier after the fix passes. These logs are archived with the runner so
that the fail-to-pass label can be checked without relying on paper-level
summaries.

\paragraph{Toolchain.}
The runners call open-source EDA tools from \texttt{PATH}, principally Verilator
and Surelog, and some load an OSS CAD Suite installation named by
\texttt{\$OSS\_CAD\_SUITE}. Requirements vary by instance, and shared helper
scripts are stored at the top level of \texttt{testbench/}.

\paragraph{Testbench-feedback disclosure.}
For the feedback condition in Table~\ref{tab:eda_feedback}, the agent receives
only the failing compile or runtime log produced by the verifier, together with
the executed command and exit status. The feedback does not include the
developer patch, the fixed-phase pass log, the gold file list, hidden testcase
rationale, or any manually written explanation of the repository.

\paragraph{Code repository.}
Each instance runs through the \texttt{run\_verilator.sh} script in its
\texttt{testbench/} directory, which
reports the buggy-phase and fixed-phase status and the fail-to-pass verdict. The
\texttt{benchmark/verifier/} directory of the code repository adds the batch
scripts used for sweeps, including \texttt{f2p\_sweep\_one.sh} and
\texttt{count\_fail\_to\_pass.py}. The \texttt{agents/} directory holds the eight
open-source agent frameworks, and \texttt{benchmark/eval\_harness/} holds
evaluation scaffolds for Gemini CLI and the feedback runs. The release does not
include scripts that regenerate
Tables~\ref{tab:product_results}, \ref{tab:agentic_study}, and
\ref{tab:eda_feedback} end to end.

\paragraph{Integrity checks.}
The Croissant metadata records a SHA-256 checksum for each of the 511 issue
files. The commit history of the data repository identifies the exact version
of the snapshots, patches, and testbenches.

\paragraph{Fail-to-pass runner example.}
The artifact also includes concrete per-instance runners; for example,
\texttt{testbench/aolofsson\_oh\_pr74/run\_verilator.sh}
documents the full fail-to-pass protocol for one general-purpose input/output (GPIO) issue in the OH repository. The runner
copies the pinned repository snapshot into a writable scratch directory, applies
the developer patch in reverse to construct the \textsc{buggy} phase, and then
executes the native smoke check plus a Verilator testbench. It then recreates
the scratch directory, applies the developer patch forward to construct the
\textsc{fixed} phase, and reruns the same checks. This instance targets
\texttt{src/common/hdl/oh\_tristate.v} and
\texttt{tb\_oh\_tristate\_verilator.sv}: before the patch, Verilator reports
\texttt{\%Error-MODMISSING} because the testbench instantiates
\texttt{oh\_tristate} but the module is absent; after the patch, the testbench
drives the bidirectional \texttt{io} port with four external values and emits
only \texttt{PASS:} lines when \texttt{in} reflects those values. The runner
accepts the instance only when the buggy phase has nonzero exit status and the
fixed phase exits zero, after which it emits patch-statistic and execution-status
log entries and reports Verilator coverage. This example shows the same pattern used by Phoenix-bench
at scale: the benchmark records the exact repository snapshot, patch, testbench,
runner command, and logs, and the resolved-rate label is derived from
deterministic EDA execution rather than manual or LLM judgment.

\section*{NeurIPS Paper Checklist}

\begin{enumerate}

\item {\bf Claims}
    \item[] Question: Do the main claims made in the abstract and introduction accurately reflect the paper's contributions and scope?
    \item[] Answer: \answerYes{}.
    \item[] Justification: The abstract and introduction state the paper's scope as a practice-driven benchmark suite, and the empirical claims are supported in Sections~\ref{sec:benchmark} and \ref{sec:eval}.

\item {\bf Limitations}
    \item[] Question: Does the paper discuss the limitations of the work performed by the authors?
    \item[] Answer: \answerYes{}.
    \item[] Justification: Appendix~\ref{app:limitations} discusses task-text quality, language and EDA-tool scope, model-version and run variability, limits of causal interpretation, and test-gaming audits.

\item {\bf Theory assumptions and proofs}
    \item[] Question: For each theoretical result, does the paper provide the full set of assumptions and a complete (and correct) proof?
    \item[] Answer: \answerNA{}.
    \item[] Justification: The paper introduces a benchmark suite and empirical evaluation, not theoretical results.

    \item {\bf Experimental result reproducibility}
    \item[] Question: Does the paper fully disclose all the information needed to reproduce the main experimental results of the paper to the extent that it affects the main claims and/or conclusions of the paper (regardless of whether the code and data are provided or not)?
    \item[] Answer: \answerYes{}.
    \item[] Justification: Sections~\ref{sec:benchmark}--\ref{sec:eval} describe the data collection funnel, verification environment, evaluated systems, models, and metrics; Appendix~\ref{app:artifact} lists the public data, runners, and code.

\item {\bf Open access to data and code}
    \item[] Question: Does the paper provide open access to the data and code, with sufficient instructions to faithfully reproduce the main experimental results, as described in supplemental material?
    \item[] Answer: \answerYes{}.
    \item[] Justification: Code is available at \url{https://github.com/zanehpn/Phoenix-Bench}; benchmark data are hosted at \url{https://huggingface.co/datasets/OvertureMMXXI/Phoenix-Bench}. Appendix~\ref{app:artifact} describes both releases, including the per-instance runners and verifier scripts that rerun fail-to-pass verification. The release does not include scripts that regenerate Tables~\ref{tab:product_results}--\ref{tab:eda_feedback} end to end.

\item {\bf Experimental setting/details}
    \item[] Question: Does the paper specify all the training and test details (e.g., data splits, hyperparameters, how they were chosen, type of optimizer) necessary to understand the results?
    \item[] Answer: \answerYes{}.
    \item[] Justification: The paper has no training procedure; the experimental section specifies benchmark instances, models, systems, EDA execution environment, and evaluation metrics.

\item {\bf Experiment statistical significance}
    \item[] Question: Does the paper report error bars suitably and correctly defined or other appropriate information about the statistical significance of the experiments?
    \item[] Answer: \answerNo{}.
    \item[] Justification: The paper reports aggregate pass/fail rates over a fixed benchmark suite without confidence intervals or repeated-run estimates. A deterministic verifier does not eliminate variability in agent-generated patches.

\item {\bf Experiments compute resources}
    \item[] Question: For each experiment, does the paper provide sufficient information on the computer resources (type of compute workers, memory, time of execution) needed to reproduce the experiments?
    \item[] Answer: \answerNo{}.
    \item[] Justification: The paper specifies the EDA tools, but it does not yet provide full wall-clock time, CPU/memory, or total compute cost for every experiment.
    
\item {\bf Code of ethics}
    \item[] Question: Does the research conducted in the paper conform, in every respect, with the NeurIPS Code of Ethics \url{https://neurips.cc/public/EthicsGuidelines}?
    \item[] Answer: \answerYes{}.
    \item[] Justification: The work studies public hardware repositories and LLM-agent evaluation without human-subject experiments or deployment on users.

\item {\bf Broader impacts}
    \item[] Question: Does the paper discuss both potential positive societal impacts and negative societal impacts of the work performed?
    \item[] Answer: \answerNo{}.
    \item[] Justification: The current draft frames the benchmark's research utility and limitations, but it does not yet include a dedicated broader-impact discussion covering both benefits and misuse risks.
    
\item {\bf Safeguards}
    \item[] Question: Does the paper describe safeguards that have been put in place for responsible release of data or models that have a high risk for misuse (e.g., pre-trained language models, image generators, or scraped datasets)?
    \item[] Answer: \answerNA{}.
    \item[] Justification: The paper does not release a pretrained model or high-risk generative asset; the benchmark is derived from public hardware repositories and does not include private user data.

\item {\bf Licenses for existing assets}
    \item[] Question: Are the creators or original owners of assets (e.g., code, data, models), used in the paper, properly credited and are the license and terms of use explicitly mentioned and properly respected?
    \item[] Answer: \answerYes{}.
    \item[] Justification: Appendix~\ref{app:artifact} describes the instance index and licensing: instance identifiers and links name the upstream repositories, Phoenix-bench's own contribution is MIT licensed, and redistributed snapshots remain under their upstream licenses.

\item {\bf New assets}
    \item[] Question: Are new assets introduced in the paper well documented and is the documentation provided alongside the assets?
    \item[] Answer: \answerYes{}.
    \item[] Justification: Phoenix-bench is documented in Sections~\ref{sec:benchmark} and \ref{sec:eval}, including construction, labels, verification protocol, and framework interfaces.

\item {\bf Crowdsourcing and research with human subjects}
    \item[] Question: For crowdsourcing experiments and research with human subjects, does the paper include the full text of instructions given to participants and screenshots, if applicable, as well as details about compensation (if any)? 
    \item[] Answer: \answerNA{}.
    \item[] Justification: The work does not involve crowdsourcing or human-subject experiments.

\item {\bf Institutional review board (IRB) approvals or equivalent for research with human subjects}
    \item[] Question: Does the paper describe potential risks incurred by study participants, whether such risks were disclosed to the subjects, and whether Institutional Review Board (IRB) approvals (or an equivalent approval/review based on the requirements of your country or institution) were obtained?
    \item[] Answer: \answerNA{}.
    \item[] Justification: The work does not involve crowdsourcing or human-subject experiments.

\item {\bf Declaration of LLM usage}
    \item[] Question: Does the paper describe the usage of LLMs if it is an important, original, or non-standard component of the core methods in this research? Note that if the LLM is used only for writing, editing, or formatting purposes and does \emph{not} impact the core methodology, scientific rigor, or originality of the research, declaration is not required.
    \item[] Answer: \answerYes{}.
    \item[] Justification: LLMs are central to the evaluated agents; the paper describes the model backbones, agent interfaces, and LLM-based components used in the study.

\end{enumerate}

\end{document}